\documentclass[showpacs,pra]{revtex4}

\begin{document}

\title{Minimal mass-size of a stable $^3$He cluster}

\author{R. Guardiola}
\altaffiliation{On leave of absence from
Departamento de F\'{\i}sica At\'omica y Nuclear,
Facultad de F\'{\i}sica, E-46100-Burjassot, Spain}
\affiliation{Institute de Recherches Subatomiques, IN2P3-CNRS/Université\'e
Louis Pasteur, F-67037 Strasbourg, France }
\author{J. Navarro}
\affiliation{
IFIC (CSIC-Universidad de Valencia),
Apartado Postal 22085, E-46071-Valencia, Spain}

\begin{abstract}
The minimal number of $^3$He atoms required to form a bound cluster has
been estimated by means of a Diffusion Monte Carlo procedure within the
fixed-node approximation. Several importance sampling wave functions have been 
employed in order to consider different shell-model configurations. The 
resulting upper bound for the minimal number is 32 atoms. 
\end{abstract}


\maketitle

Helium clusters are unique systems for studying the properties of finite
sized quantum objects. The first systematic microscopic calculation of their 
ground state properties was carried out by Pandharipande and 
coworkers~\cite{pandha86} eighteen years ago. One interesting finding is that, 
contrary to the $^4$He case, a minimum number of $^3$He atoms is required to 
create a stable cluster, as a consequence of the large zero-point motion and 
the Pauli effect. The calculations of Ref.~\cite{pandha86} were based on an 
optimized trial wave function, where the fermions are assumed to fill harmonic 
oscillator single-particle states $1s^2 1p^6 1d^{10} 2s^2 1f^{14} 2p^6 \dots$ 
The outcome was that twenty atoms are not enough to form a bound system, but 
forty atoms are bound, and in between a {\em critical} or {\em threshold} 
number exists for having stable clusters. From the experimental point of
view, there is only indirect evidence of such a critical number, since the
smaller $^3$He systems detected insofar contains thousand of atoms.  

The critical number was calculated in Ref.~\cite{barr97}, by using a 
non-local finite-range density functional approach. The usual Kohn-Sham 
procedure provided the residual interaction required to perform a 
configuration interaction calculation. The obtained critical number was 29
atoms. Afterwards, we carried out~\cite{guar00,guar00a} a variational 
microscopic study based on the two-body He-He interaction known as HFD-B(He) 
interaction~\cite{HFD-B}, which reproduces accurately the properties of both 
$^4$He and $^3$He liquid. The trial wave function contained a 
Jastrow-correlated part and a self-adjustable translationally invariant 
configuration interaction part, including up to three particle-hole excitations.
Due to the variational character of the computational procedure, only an upper 
bound was obtained, and we concluded that the critical number is less or
equal to 34 atoms. 

In order to improve this estimate we have carried out several calculations 
within the fixed-node Diffusion Monte Carlo (DMC) 
procedure~\cite{Reyn82,Mosk82}, for selected systems near the previously 
determined critical number. The importance-sampling wave functions have been 
constructed in a similar way to the variational forms previously used in 
Refs.~\cite{guar00,guar00a}, but with some modifications. First of all, 
the confining part of the two-body Jastrow correlation used here has an 
exponential shape, instead of a gaussian one. Because of the longer range of 
the exponential shape one may expect to be more appropriate for systems near 
the binding threshold. Apart from this, the same backflow correlation was 
used, but the configuration interaction part has not been included, with the 
objetive of having a fast difussion Monte Carlo algorithm. 

The other relevant modification is related to the antisymmetric part
of the wave function. The fermionic antisymmetry is considered by means of 
the product of two Slater determinants, one for each spin orientation. In the 
present calculation we have assumed several shell-model orderings for these 
determinantal parts, and not simply the ordering related to the 
harmonic-oscillator single-particle potential previously considered. Actually, 
nothing is known about the ordering of shells. The only indirect information 
comes from the study of a single $^3$He atom tied to a core of $^4$He atoms, 
where calculations indicate an order based on the orbital angular momentum, 
namely $1s\ 1p\ 1d\ 1f\ \dots$, with probably the $2s$ level being interleaved 
between the $n=1$ levels~\cite{nava04,fant04}. Therefore, we have considered 
configurations with a fixed occupancy $1s^2\ 1p^6\ 1d^{10}$, the $N=18$ major 
shell, and several distributions of the remaining particles between the $2s$ 
and the $1f$ shell, which are classified by the value of the total spin $S$.

The actual calculations have been carried out with a time slice 
$\tau=0.00025 \ {\rm K}^{-1}$, with an average population of 1000 walkers and 
for 10000 time steps, plus 1000 previous steps in order to reach the stability. 
To obtain an estimate of the variance, the sampling steps were grouped  
into blocks of 100 moves.

\begin{table}[ht!]
\caption{Ground state energies $E$ (in K) obtained for several clusters near 
the critical binding number. For a given number $N$ of atoms, the shell-model 
configuration and the value of the total spin $S$ are indicated.}
\label{results}
\begin{tabular}{ccccc}
\hline
 $N$ & Config.      &  $S$ & $E \ (K)$\\
 \hline
 31  & $ 2s^2 1f^{11}$ &  3/2& Unbound\\
 31  & $ 2s^2 1f^{11}$ &  1/2& Unbound\\
 \hline
 32  & $ 2s^2 1f^{12}$ &  0  & $-0.27 \pm 0.03 $        \\
 32  & $ 2s^2 1f^{12}$ &  1  & $-0.23 \pm 0.03 $         \\
 32  & $ 2s^0 1f^{14}$ &  0  & Unbound \\
 \hline
 33  & $ 2s^1 1f^{14}$ & 1/2 & Unbound \\
 33  & $ 2s^2 1f^{13}$ & 1/2 & $-0.86 \pm 0.04$\\
 \hline
 34  & $ 2s^2 1f^{14}$ &   0 & $-1.52 \pm 0.04$\\
 \hline
\end{tabular}
\end{table}

The obtained results are presented in Table~\ref{results}. The $N=34$ cluster 
appears clearly bound, as well as one of the configurations chosen for $N=33$. 
This last result has the interest of suggesting the $2s$ shell to be deeper 
bound than the $1f$ shell. 

Regarding the $N=32$ cluster we had an initial guess in favor of the 
configuration $1s^0\ 1f^{14}$, because it would correspond to a closed shell. 
However the actual calculations show that this configuration does not result 
in a bound state, preferring instead to complete the $2s$ shell. This is 
not surprising once it has been established the shell ordering in the 
$N=33$ case. Note that the total spin of the configuration $2s^2 1f^{12}$ can 
have the two values $S=0$ and 1. Taking into account the statistical errors, 
one may conclude that these states are bound with the same binding energy. 
Either these states are degenerate or the difference in energy is smaller than 
our statistical errors, i.e. of the order of ten mK.
Given that the obtained energies are very close to zero, the imaginary time 
evolution was carried out for these two cases for as much as 40000 time steps. 

Finally we have considered the $N=31$ cluster, with two spin states ($S=1/2$ 
and $S=3/2$) of the $2s^2 1f^{11}$ configuration. It turns out that neither of 
these states is bound, their energy being close to zero but positive.

In conclusion, we have find $N=32$ as an upper bound to the minimal 
mass-size of a stable $^3$He cluster. The various configurations here considered
correspond in practice to probe different nodal surfaces. It is worth 
stressing that our results indicate that, within the computational statistical 
errors, the binding energy is independent of the spin coupling, depending only 
on the chosen configuration. Finally, it should also be mentioned that the 
separation energy for $N=34$ ($0.66 \pm0.06$ K) is almost the same as that of 
$N=33$ ($0.59\pm 0.05$, $0.64 \pm 0.05$ K), thus suggesting a single-particle 
structure of these fermionic clusters, with a residual interaction compatible 
with zero. The determination of the precise critical number should require a 
calculation beyond the variational fixed-node approximation, and it is not 
excluded that the result could depend on the He-He interaction employed in the 
practical calculation.

\acknowledgments
This work has been supported by MCyT/FEDER (Spain), grant number BMF2001-0262 
and GV (Spain), grant number GV2003-002.  One of us (RG) acknowledges the 
IReS (Strasbourg) by his hospitality.

\end{document}